\documentclass[aps,prl,preprint,groupedaddress]{revtex4}

\usepackage{epsfig}
\usepackage{amssymb}
\usepackage{times}

\begin{document}

\def\lambar{\lambda \hspace*{-5pt}{\rule [5pt]{4pt}{0.3pt}} \hspace*{1pt}}

{\large  DESY 21-162}

{\large  October 2021}

\bigskip
\bigskip
\bigskip
\bigskip
\bigskip
\bigskip


\title{Application of a modified chirp-taper scheme for generation of attosecond pulses in XUV and soft X-ray FELs}



\author{E.A.~Schneidmiller}
\email[]{evgeny.schneidmiller@desy.de}
\affiliation{Deutsches Elektronen-Synchrotron DESY,
Notkestr. 85, 22607 Hamburg}

\date{\today}

\begin{abstract}
Typically, in Self-Amplified Spontaneous Emission Free Electron Laser (SASE FEL) based short-pulse schemes, pulse duration is limited by FEL coherence time. For hard X-ray FELs, coherence time is in a few hundred attosecond range while for XUV and soft X-ray FELs it is in the femtosecond regime. In this paper the modification of so-called chirp-taper scheme is developed that allows to overcome the coherence time barrier. Numerical simulations for XUV and soft X-ray FEL user facility FLASH demonstrate that one can generate a few hundred attosecond long pulses in the wavelength range 2 - 10 nm with peak power reaching hundreds of megawatts. With several thousand pulses per second this can be a unique source for attosecond science.
\end{abstract}


\maketitle

\section{Introduction}

Attosecond science \cite{attoscience} is rapidly developing nowadays thanks to the laser-based techniques such as 
chirped-pulse amplification and high-harmonic generation. 
There are also different schemes proposed for generation of attosecond X-ray pulses in free electron lasers
\cite{attofel-oc,oc-2004-2,atto-b,atto-e,atto-f,prstab-2006-2,ding,huang-24as}.
Many of these schemes make
use of a few-cycle intense laser pulse to modulate electron energy in a short undulator, and then to make only a short slice (a fraction of wavelength) efficiently lase in a SASE undulator. In particular, in the chirp-taper scheme \cite{prstab-2006-2}, a slice with the strongest energy chirp is selected for lasing by application of a strong reverse undulator taper that compensates FEL gain degradation within that slice. The lasing in the rest of the bunch is strongly suppressed due to uncompensated reverse taper.

Creation of a short lasing slice can also be done without using a laser. In particular, nonlinear compression of multi-GeV electron beams \cite{shuang} and self-modulation in a wiggler of a bunch with the special temporal shape \cite{duris} allowed to generate a few hundred attosecond long pulses at the Linac Coherent Light Source (LCLS). 
However, creation of sub-femtosecond features in the electron bunch at lower electron energies ($\simeq 1$ GeV) is problematic.

Typically, pulse duration in SASE-based short-pulse schemes is limited by FEL coherence time \cite{book}. For hard X-ray FELs, coherence time is usually in a few hundred attosecond range. For such a case an adequate choice of a laser could be a Ti:Sapphire system providing a few mJ within 5 fs (FWHM) with the central wavelength at 800 nm. However, for XUV and soft X-ray regimes the coherence time is in femtosecond range, and a longer wavelength laser is needed \cite{fawley} to match a lasing slice duration and coherence time. 
In this paper a simple method is developed \footnote{A similar concept was proposed by the author as an option for FLASH upgrade \cite{FL-CDR} but was not studied.} 
to overcome this barrier and to produce XUV and soft X-ray pulses that are much shorter than FEL coherence time, and can be as short as few hundred attoseconds.

\section{Principles of operation}

\begin{figure}[tb]

\includegraphics[width=1.0\textwidth]{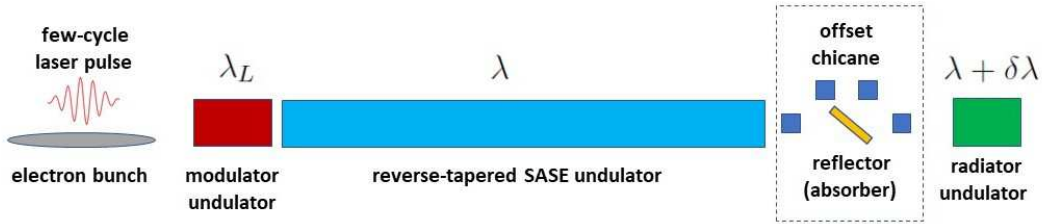}

\caption{
Conceptual scheme for generation of attosecond pulses. Dashed rectangle illustrates a particular realization of suppression (separation) of a radiation background from the SASE undulator.
}
\label{fig:atto-scheme}
\end{figure}

\begin{figure}[tb]

\includegraphics[width=0.6\textwidth]{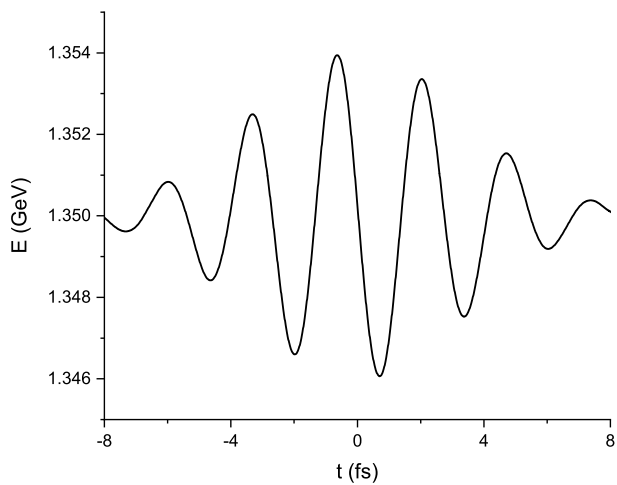}

\caption{
Energy modulation induced by the laser. Bunch head is on the left side.
}
\label{fig:mod4MeV}
\end{figure}

Conceptual representation of the attosecond scheme is shown in Fig.~\ref{fig:atto-scheme}. Few-cycle laser pulse is used to modulate a central part of an electron bunch in energy in a short (typically, two-period) modulator undulator. The wavelength $\lambda_L$ is chosen such that the lasing slice is much shorter than FEL coherence length. In particular, for generation of attosecond pulses in XUV and soft X-ray regime one can consider Ti:sapphire laser. A typical shape of energy modulation after the modulator undulator is shown in Fig.~\ref{fig:mod4MeV}. 

Then the bunch enters a long SASE undulator tuned to a wavelength $\lambda$. The undulator is operated in the same way as in the classical chirp-taper scheme \cite{prstab-2006-2}: it is reverse-tapered to compensate for the energy chirp within the central slice (positioned at $t=0$ in Fig.~\ref{fig:mod4MeV}). In this way the FEL gain degradation within this slice is avoided, and the amplification proceeds up to the onset of saturation. The rest of the bunch suffers from the uncompensated reverse taper, and the lasing is strongly suppressed (except maybe for two satellites positioned around $t = \pm 2.7$ fs on Fig.~\ref{fig:mod4MeV} with the negative time derivative). The difference with the standard scheme is that now the central lasing slice is much shorter than FEL coherence time. The distribution of bunching (density modulation amplitude) is rather narrow and is localized at the end of that slice but the radiation slips forward, and a relatively long pulse (on the order of coherence time) is produced. 

The next task is to get rid of this relatively long radiation pulse (as well as of the background radiation from the rest of the bunch) while preserving the bunching. This can be done in different ways. In Fig.~\ref{fig:atto-scheme} a possible realization is illustrated: an offset chicane with a reflector or absorber inside. Alternative options are discussed below in this Section: excessive reverse taper, an achromatic bend, a kick with a quadrupole, a dogleg, and a harmonic afterburner. 

Finally, the microbunched beam radiates in a short radiator undulator. The bunching is strong in the central slice, it is weaker in the two satellites around $t = \pm 2.7$ fs, and much weaker in the rest of the bunch. Note that reverse tapering is very efficient in suppression of the radiation but the bunching can reach high values, depending on conditions \cite{rev-tap}. We use sufficiently strong chirp and taper to make sure that the the bunching stays at low level in the whole bunch except for the mentioned slices. In addition to that, another feature of the process is used to strongly suppress the ratiation from unwanted parts of the bunch, including satellites. Namely, the central slice is stretched in the main undulator due to a strong energy chirp so that the frequency of the bunching is red-shifted with respect to the resonance frequency at the entrance of SASE undulator. The satellites have a weaker red shift. The rest of the bunch also has a red shift due to undulator taper \cite{stupakov-taper} but it is even weaker. The radiator undulator is set to the resonance with the central slice, and the number of periods is approximately equal to the number of cycles of density modulation within that slice. The other parts of the bunch are non-resonant and radiate very weakly. Below in this Section the operation of the radiator is discussed in more details. 

As a result, few hundred attosecond long pulses with low background can be produced in XUV and soft X-ray ranges.
The prerequisite for operation of this scheme is a sufficiently long SASE undulator. Note that there is always the range of photon energies at XUV and X-ray FEL facilities for which the saturation occurs well before the undulator end, and there is a reserve for operation with different advanced schemes.

\subsection{Chirp-taper compensation effect}

If there is a linear energy chirp at the undulator entrance, it can have a significant  effect on SASE FEL properties, in particular on the gain. The strength of this effect can be characterized by the energy chirp parameter \cite{prstab-2006-2}:

\begin{equation}
\hat{\alpha} = -\frac{d \gamma}{dt} \frac{1}{\gamma_0 \omega_0 \rho^2} \ ,
\label{eq:chirp-parameter}
\end{equation}

\noindent where $\rho$ is the well-known FEL parameter \cite{bon-rho, book}, and $\gamma$ is relativistic factor.
Factor $\gamma_0$ for a reference particle 
and reference frequency $\omega_0$ are connected by the FEL resonance
condition: $\omega_0 = 2ck_w \gamma_0^2/(1+K^2/2)$. Here $K$ is the undulator parameter and $k_w = 2\pi/\lambda_w$ with $\lambda_w$ being the undulator period.

It was shown in \cite{prstab-2006-2} that a degrading effect of a linear energy chirp on SASE FEL gain can be compensated for by applying a linear undulator taper as soon as the following condition is satisfied:

\begin{equation}
\frac{d K}{dz} = - \frac{(1+K_0^2/2)^2}{K_0} \frac{1}{\gamma_0^3} \frac{d \gamma}{c dt}
\label{compensation}
\end{equation}

\noindent Here $K_0$ is the value of undulator parameter at the undulator entrance. Note that the condition (\ref{compensation}) is applicable when $4 \pi \rho \hat{\alpha} \ll 1$, and a perfect compensation is only possible in the limit $\rho \to 0$. However, for practical applications a perfect compensation is usually not required.

\subsection{Chirp-taper compensation for a short lasing slice}

Operation of SASE FELs with short bunches was studied in \cite{bon-short,we-short}. A relevant parameter to characterize an effect of bunch length on FEL operation is $\rho \omega \sigma_z/c$ with $\sigma_z$ being rms bunch length. When this parameter is smaller than one (i.e. when bunch is shorter than FEL coherence length $c(\rho \omega)^{-1}$), one can observe an increase of saturation length and a reduction of FEL efficiency.

The condition (\ref{compensation}) is also valid for short bunches or for short lasing slices (as it was mentioned above, an ideal compensation is only possible when $\rho \to 0$). In this paper we deal with long bunches but short lasing slices having the strongest laser-induced energy chirp.
For such a case, instead of $\sigma_z$ one can consider a reduced laser wavelength, $\lambar_L = \lambda_L/(2\pi)$. Thus, a relevant parameter is now $\rho \lambda_L/\lambda$ where 
$\lambda = 2 \pi c/\omega$ is the FEL wavelength. It follows from numerical simulations with laser-modulated beam that an increase of saturation length for a short lasing slice with respect to a normal SASE with long bunches can be approximated as follows:

\begin{equation}
\frac{L_{\mathrm{sat}}}{L_{\mathrm{sat}}^{\mathrm{(long \ bunch)}}} \simeq \left(\rho \frac{\lambda_L}{\lambda} \right)^{-1/2}  \ \ \ \ \ \ \ \mathrm{for} \ \ \ \ \ \ \  \rho \frac{\lambda_L}{\lambda} < 1
\label{short-sat}
\end{equation}

The dependence is similar to that for short bunches \cite{bon-short,we-short}.
For the purpose of the proposed scheme, one should stop at the onset of saturation (typically 80\% to 90\% of saturation length) to avoid an increase of the width of bunching distribution within the lasing slice.
Thus, a total increase of the required undulator length can be acceptable in many practical cases even for a small parameter $\rho \lambda_L/\lambda$.

\subsection{Suppression of background from the main undulator}

One of the advantages of the proposed scheme is that one can get a clean attosecond pulse from the afterburner. However, we need to get rid of the background produced in the main undulator. Let us consider possible ways of doing this.

\subsubsection{Excessive reverse taper}

Reverse taper is efficient in suppression of the radiation, although under some conditions the bunching can survive \cite{rev-tap}. In case of the considered scheme one can apply reverse taper which is stronger than the one needed for compensation of the energy chirp in the main lasing slice. With some delay of the saturation, one can get a strong bunching there but almost no radiation. Excess of reverse taper would then be even stronger in the adjacent slices with the same sign of energy chirp but a weaker amplitude. There one can suppress the radiation even stronger, and the bunching factor saturation is delayed also stronger than in the main slice. In general, the intensity of the radiation from the main undulator can be made sufficiently small. Then, in the radiator a strong power is produced only within the main slice, also due to a frequency offset mentioned above \footnote{Note that in some cases a regular undulator segment (if it is sufficiently short) with optimized K value can play a role of the radiator undulator.}. 
A disadvantage of this method is that it requires a longer main undulator which is not always possible. Also, bunching within the main slice can be weaker than in a case without over-compensation, depending on parameters.

\subsubsection{Achromatic bend or a kick with a quadrupole}

Another way to produce clean attosecond pulses in the afterburner is to create an angle between the radiation from the main SASE undulator and from the radiator by using an achromatic bend \cite{bend-kulipanov} or 
a kick with a quadrupole \cite{macarthur}. The latter technique (in combination with reverse taper) was successfully used for generation of circularly polarized radiation with high purity at LCLS \cite{lutman-circ}. 

\subsubsection{Chicane or dogleg}

One can also create an offset between the electron beam and the radiation from the SASE undulator with the help of a chicane, as shown in Fig.~\ref{fig:atto-scheme}. Then the radiation is either absorbed directly or reflected to an absorber. A possible difficulty is that the longitudinal dispersion, characterized by a transfer matrix coefficient $R_{56}$, is generated in the chicane. This can be a useful effect: additional bunching can be created, so that one can stop earlier in SASE undulator; moreover the lasing slice is stretched even stronger which helps in suppression of background in the radiator. 
However, these two functions of the chicane (a technically reasonable offset and an optimal $R_{56}$) should be matched which is not always easy to do. A more flexible system could be a chicane with quadrupoles in the dispersion regions \cite{thompson} so that one can efficiently control $R_{56}$ while the required offset is kept.

Another possible solution is a dogleg that creates a sufficient offset but the $R_{56}$ is typically too small to influence longitudinal dynamics.

\subsubsection{Harmonic afterburner}

Radiation at the even harmonics of SASE undulator is weak. Thus, tuning the radiator to the second harmonic, for example, would help to provide low-background attosecond pulses. Radiation at the fundamental of the undulator can be filtered out if it disturbs an experiment.

\subsection{Suppression of satellites in the afterburner}

One of the problems of laser-based methods for production of the attosecond pulses is an insufficient contrast of laser modulation that leads to generation of satellite pulses shifted in time by a cycle of the laser light \cite{atto-f}. They are weaker than the main pulse but can still be a problem for user experiments. In the proposed scheme we rely not only on a less efficient generation of bunching for the satellites, but also (and mainly) on the fact the frequency of bunching in the main slice is different (more red-shifted) from that in the satellites. The radiator is tuned to the frequency of the main slice, and the radiation from the adjacent slices is strongly suppressed because of the offset from resonance. Spectral properties of the radiator are characterized by the well-known sinc-function:

\begin{equation}
f_1 (\omega) = \left( \frac{\sin (N_w \pi \frac{\omega -  \omega_r}{\omega_r})}{N_w \pi \frac{\omega - \omega_r}{\omega_r}} \right)^2
\label{sinc}
\end{equation}

\noindent Here $\omega_r$ is the resonance frequency of the radiator and $N_w$ is the number of periods. The latter parameter should be chosen such that it is approximately the same as the number of cycles in the bunching distribution within the main lasing slice. At the same time, as it can be seen from (\ref{sinc}), for an efficient suppression one needs to satisfy the condition $N_w \ge \omega_m/(\omega_s - \omega_m)$ with $\omega_m$ being the frequency of bunching in the main slice and $\omega_s$ in the satellites. One can even adjust parameters such that the satellites are positioned in frequency domain at the zeros of the sinc function, i.e. when $N_w \simeq n \omega_m/(\omega_s - \omega_m)$, where $n$ is a natural number. In this case the suppression will be especially effective.

The density modulations in the bulk of the beam (not modulated by the laser) are much weaker than those on the slopes. In addition, they have a much larger frequency offset from the resonance in the radiator, so that the radiation is strongly suppressed. As a result, one can obtain a clean attosecond pulse from the radiator.

\section{Numerical simulations for FLASH}

In the first XUV and soft X-ray FEL user facility
FLASH \cite{flash-nat-phot,njp} the electron bunches with maximum energy of 1.25 GeV are distributed between the two
undulator lines. The facility operates in the wavelength range 4 - 60 nm with long pulse trains (several hundred pulses) following with 10 Hz repetition rate. After the planned upgrade, the electron energy will reach 1.35 GeV, this energy is used in numerical simulations. 
Electron bunches with the charge 100 pC and the following quality \cite{zemella} are considered in this paper: peak current 1.5 kA, normalized emittance 0.5 mm mrad, uncorrelated energy spread 200 keV. Parameters of the second undulator line FLASH2 are used in the simulations. Segmented variable-gap undulator with the period of 3.14 cm and the maximum K about 2.7 consists of twelve 2.5 m long segments with quadrupoles in the intersections. Average beta-function of the FODO structure is 7 m.

Modulator undulator has two periods with the period length of 15 cm and the K value of 12. The Ti:Sapphire laser system generates 5 fs long pulses (FWHM intensity), a pulse energy is 0.25 mJ. The Rayleigh length is chosen to be 1 m. Energy modulation of the electron beam for this parameter set of laser-modulator system is presented in Fig.~\ref{fig:mod4MeV}, the maximum energy deviation is 4 MeV.

FEL simulations were performed with the code SIMPLEX \cite{simplex}.
The results of simulations at three different wavelengths with respectively optimized afterburners are presented below for illustration.

\subsection{Case I: the fundamental at 4.7 nm}

\begin{figure}[tb]

\includegraphics[width=0.5\textwidth]{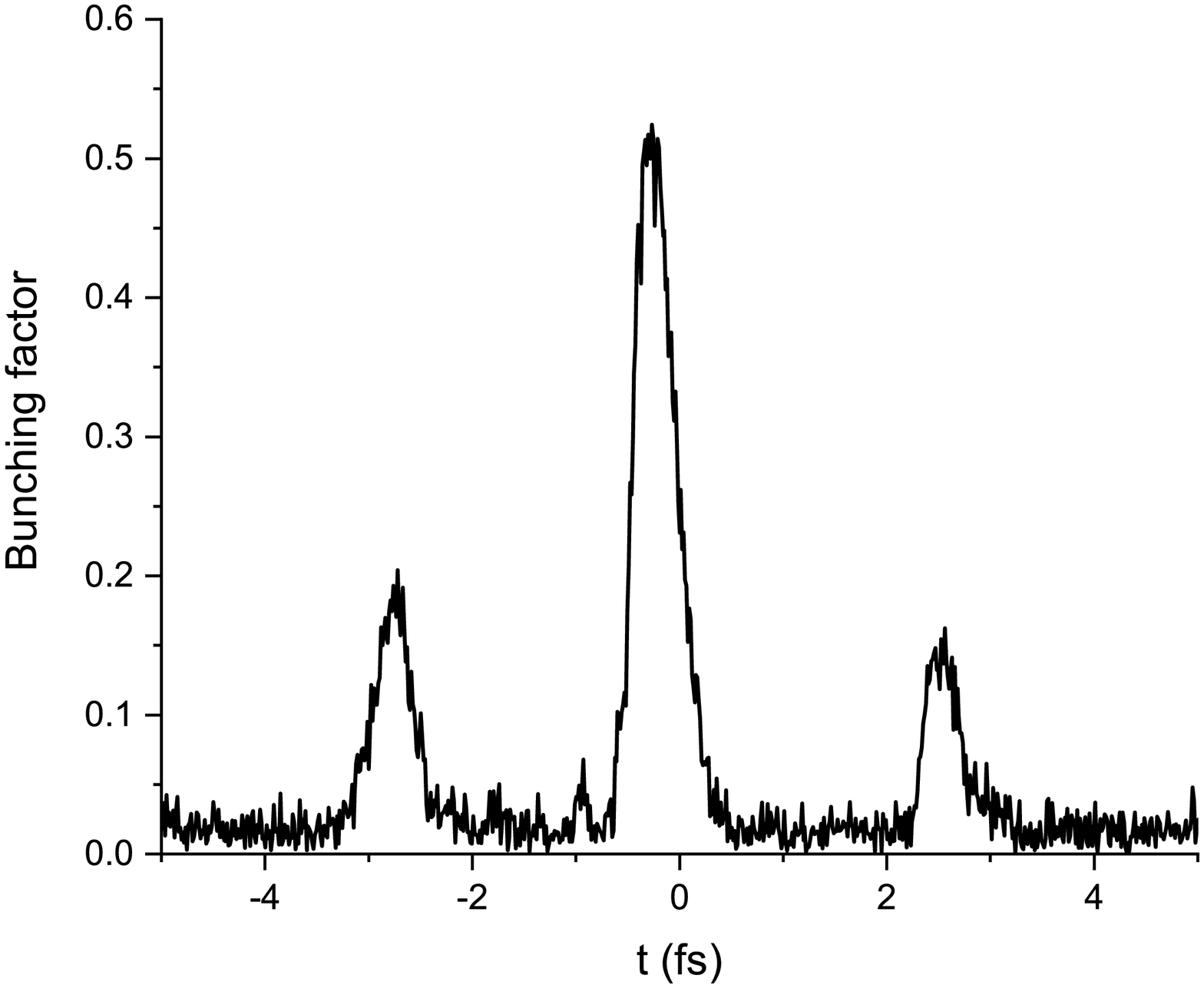}
\includegraphics[width=0.5\textwidth]{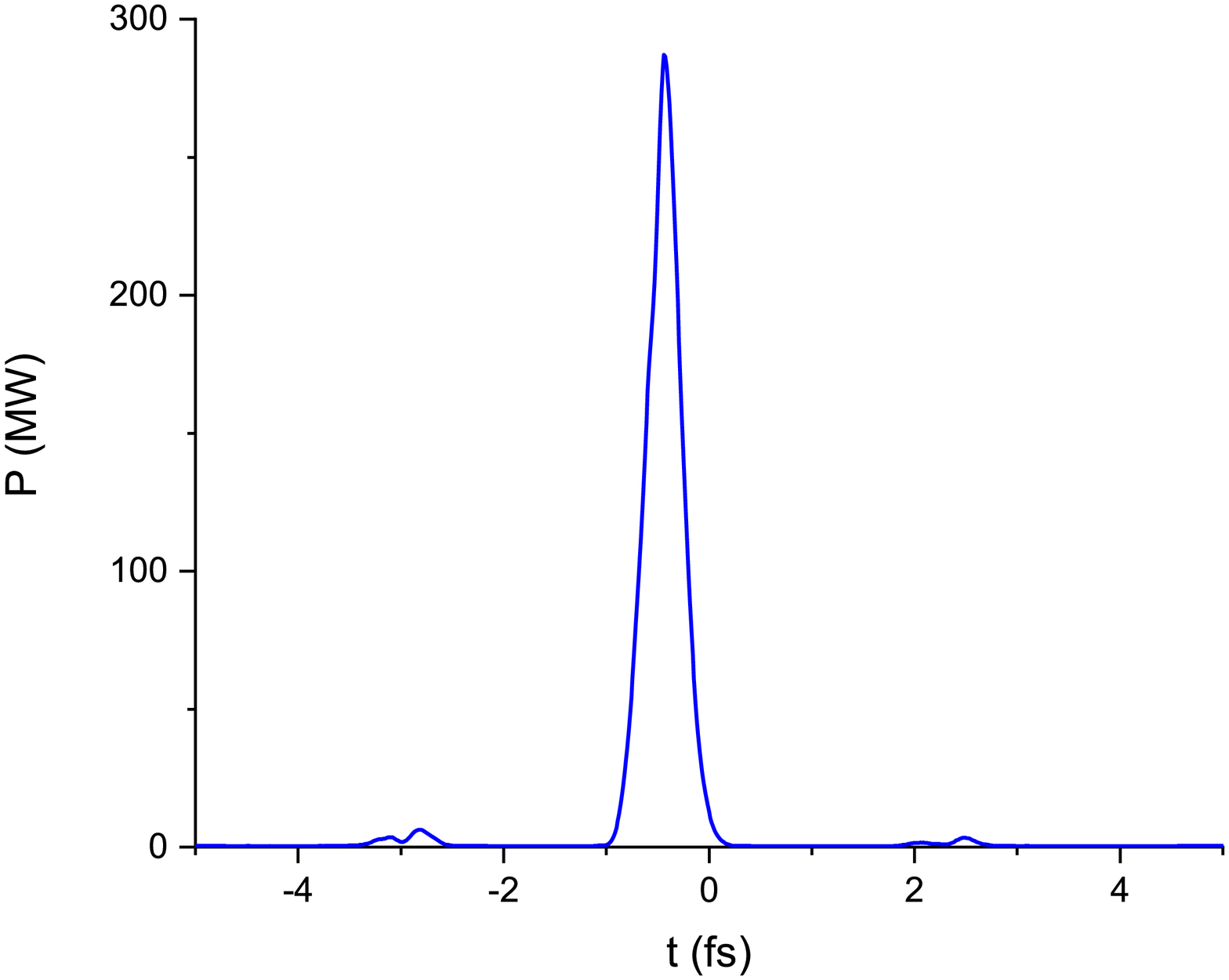}

\caption{
Bunching factor at the entrance to the radiator (upper plot) and power at its exit (lower plot) for a single shot. Bunch head is on the left side.  
}
\label{fig:4nm}
\end{figure}

Let us first consider the case when energy-modulated beam (see Fig.~\ref{fig:mod4MeV}) radiates in FLASH2 undulator tuned to 4 nm at the entrance ($K_0 = 1.25$). Since the parameter $\rho$ for the considered beam and undulator parameters is $1.9\times 10^{-3}$, one can use (\ref{eq:chirp-parameter}) to find that $\hat{\alpha} \simeq 4$ for the central slice.
The reverse step-taper is applied such that $K$ increases by 0.025 in each undulator segment, we use ten segments in simulations. Note that the chosen reverse taper is about $20 \%$ stronger than the one needed for the perfect compensation. This helps reduce background from the main undulator without affecting significantly the generation of strong bunching within the central slice. However, a much stronger excessive reverse taper would lead to a significant increase of the undulator length and cannot be considered as the main method of background reduction for given parameters. 

Let us discuss an increase of undulator length with respect to that needed for saturation of long bunch with the same slice parameters (which is 18 m). The parameter $\rho \lambda_L/\lambda$ is 0.38 in the considered case, so that an increase of saturation length is about 60 \% according to Eq.~(\ref{short-sat}). We do not aim at reaching saturation since there is a broadening of the bunching distribution at that point. Lasing is stopped a bit earlier, at about 90 \% of the saturation length, so that the required increase of the undulator length is about 40 \% (from 18 m to 25 m).

The distribution of bunching factor in the modulated part of the bunch at the exit of the tenth undulator segment is shown for a single shot in Fig.~\ref{fig:4nm} (upper plot). One can see not only strong bunching within the central lasing slice but also a significant bunching in the satellites. In this simulation we assume that the $R_{56}$ between the main undulator and the radiator is negligible (to avoid an effect on bunching distribution it should be below $\simeq 1 {\mu}m$). Thus, the electron beam without modifications is sent to the radiator while the radiation from the main undulator is suppressed with the help of one of the methods discussed in the previous Section. 

The radiator is the short undulator with 40 periods, period length of 2.5 cm and the undulator parameter 1.804. In Fig.~\ref{fig:4nm} (lower plot) one can see the temporal profile of radiation pulse at 4.7 nm emitted by the beam with bunching shown in Fig.~\ref{fig:4nm} (upper plot). The wavelength increase is due to stretching of the central slice in the main undulator. One can also see that satellites are strongly suppressed despite a significant bunching factor, the mechanism is explained above. Total background (that includes satellites and the radiation produced in the bulk of the beam) does not exceed a few per cent level.

\begin{figure}[tb]

\includegraphics[width=0.8\textwidth]{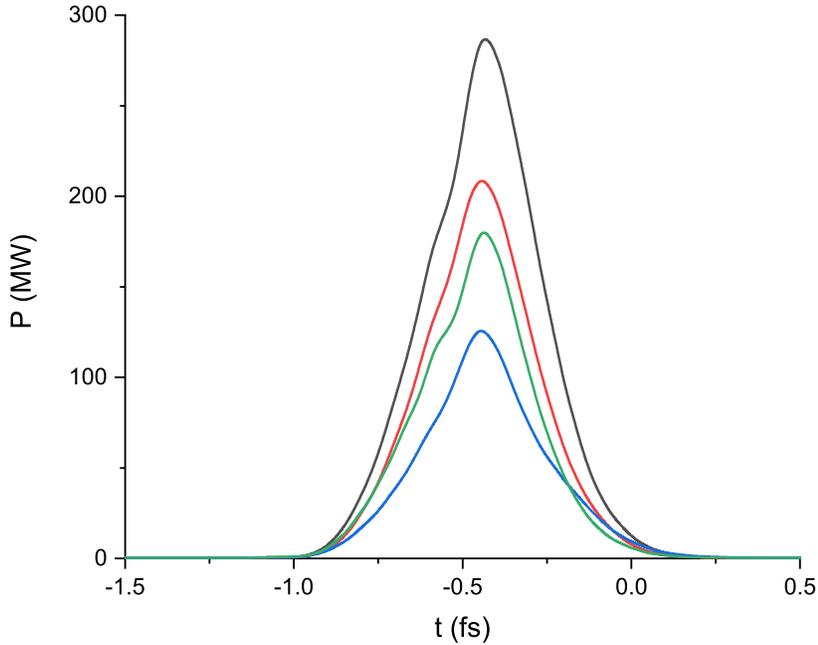}

\caption{
Radiation power for four representative shots at 4.7 nm. Bunch head is on the left side.
}
\label{fig:4nm-four-shots}
\end{figure}

Forty simulation runs were performed to study the properties of the attosecond pulses. Four representative shots are shown for illustration in Fig.~\ref{fig:4nm-four-shots}. Pulse duration is in the range 300 - 400 as (FWHM) which is by an order of magnitude smaller than FEL coherence time in the main undulator. The average pulse energy is 70 nJ with the rms shot-to-shot variations about 40 \%. It is interesting to note that despite a significant variation of pulse energy, the timing is very stable (contrary to the case of a standard single-mode lasing of short electron bunches \cite{duesterer}). This opens up a possibility of pump-probe experiments with near infrared (NIR) and soft X-ray pulses keeping sub-cycle synchronization. Indeed, the laser-modulated beam after emission of an attosecond X-ray pulse can be sent through magnetic chicane for conversion of energy modulation into density modulation on the scale of laser wavelength. Then it can radiate a NIR pulse in the short undulator (similar to modulator undulator).
Both pulses (soft X-ray and NIR) can be transported to a user instrument through the same mirror system thus preserving timing. At the experiment one can either use NIR pulse directly or (if it is too weak) do cross-correlation with a powerful laser pulse thus preserving timing information for every shot.  

Finally, let us discuss accuracy of the predictions with FEL simulation codes in the considered case of the strong energy chirp. Note that the parameter $4 \pi \rho \hat{\alpha}$ is about 0.1 which is sufficiently small for FEL theory with energy-chirped beams to be applicable. At the same time, one cannot expect a per cent level accuracy of the predictions, they are rather in the ten per cent range. In particular, one can expect a significant stretching of the central lasing slice in that range. The evolution of the process in time-frequency domain is correctly simulated by FEL codes (all the necessary information is contained in phases), however the change of average electron density is neglected in these codes. Thus, the radiated power is somewhat overestimated in the simulations presented here.

\subsection{Case II: the second harmonic at 2.4 nm}

\begin{figure}[tb]

\includegraphics[width=0.8\textwidth]{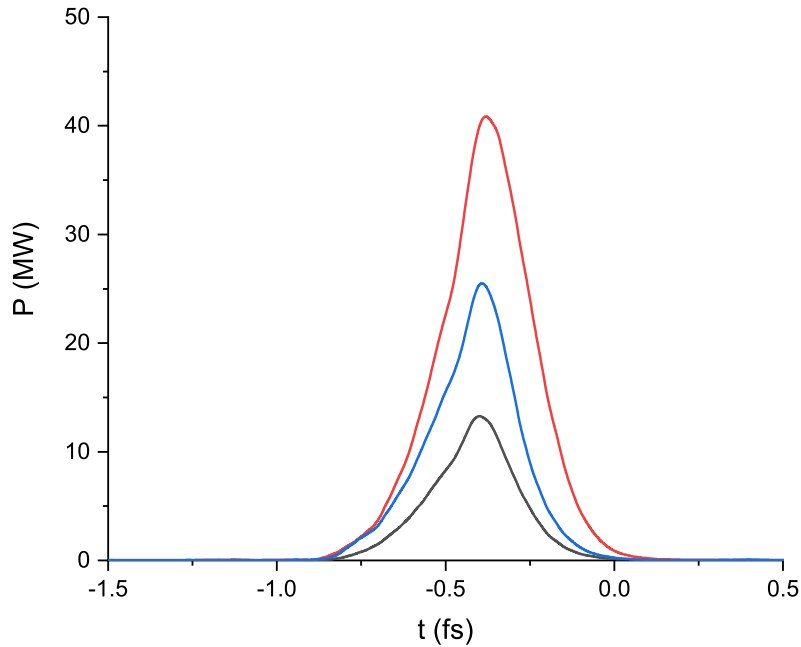}

\caption{
Radiation power for three representative shots at 2.4 nm. Bunch head is on the left side.
}
\label{fig:2nm}
\end{figure}

The macroparticle distributions from simulation runs at 4 nm in the main undulator were used for simulations of the radiation at the second harmonic in the dedicated afterburner. The undulator  parameters are as follows: period length is 2 cm, number of periods is 20, K equals 1.134. Three representative shots are shown in Fig.~\ref{fig:2nm}. Again, due to the stretching, the wavelength of the second harmonic is not 2 nm but 2.4 nm. Pulse durations are in the range 250 - 300 as (FWHM), and the average pulse energy is 6 nJ.

\subsection{Case III: the fundamental at 9.3 nm}

\begin{figure}[tb]

\includegraphics[width=0.8\textwidth]{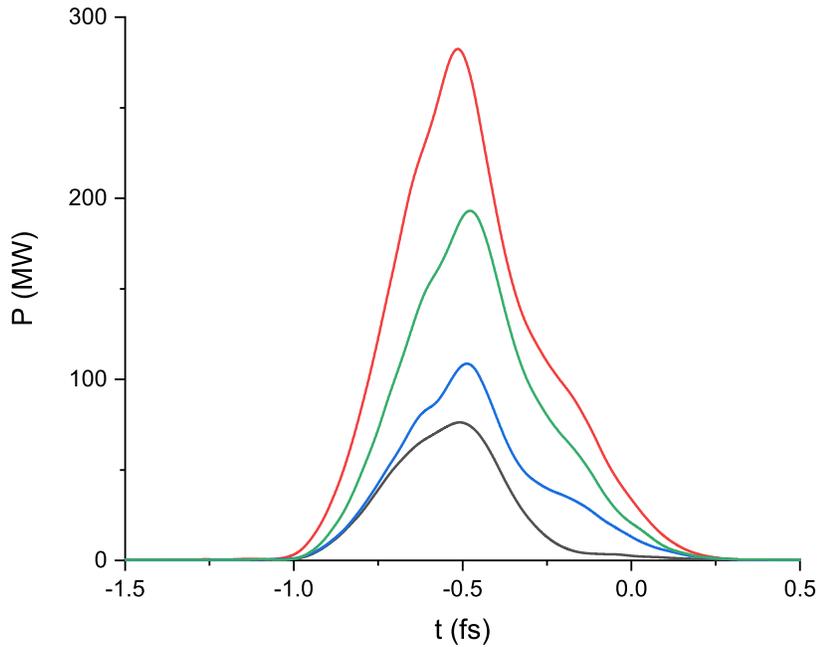}

\caption{
Radiation power for four representative shots at 9.3 nm. Bunch head is on the left side.
}
\label{fig:9nm}
\end{figure}

It is also constructive to illustrate the operation of the scheme at a longer wavelength. The electron beam is the same as it was in the previous simulations. The main undulator is now shorter, it consists of nine segments. The period of the afterburner is the same as that of the main undulator, but the number of periods is only 25, so that it is shorter than a segment of the main undulator. 
Four representative shots are shown in Fig.~\ref{fig:9nm}. Average pulse energy is 75 nJ, and the pulse duration is about 400 as (FWHM).  

Note that three different periods of the afterburners were used in three considered cases. They are all relatively short (about 1 m) and can be placed behind each other. In practice one can optimize the parameters and the number of devices depending on the operating range of the attosecond facility and a range of electron energy (in simulations only one energy was used).   

\section{Prospects for hard X-rays}

The standard chirp-taper attosecond scheme was originally intended for application in hard X-ray FELs. Indeed, the coherence time (a few hundred attoseconds) naturally matches lasing slice when a few-cycle short wavelength laser (like Ti:sapphire) is used. The scheme, described in this paper, can potentially be applied to push pulse duration to few tens attoseconds regime if the following approach is used. One can apply so-called eSASE scheme \cite{atto-f} and use the same laser to create energy modulation in the electron bunch with a subsequent conversion of these modulations into density modulations (current spikes). The duration of the central spike with the highest peak current is then in sub-hundred attosecond range.
A strong energy chirp is accumulated along this spike due to longitudinal space charge field in front and inside of the SASE undulator \cite{sc-und}. Then the chirp-taper compensation is used in the same way as in the case of laser-induced chirp. And to avoid pulse lengthening due to the slippage one can use the method developed in this paper. Note that in some cases the application of excessive reverse taper can be adequate, and a regular segment of an X-ray FEL undulator can be used as a radiator. One can also consider the second harmonic generation in one segment. In other words, application of this method might not require any modification of existing SASE undulators.

\section{Conclusion}

A modification of the chirp-taper scheme allows to produce FEL pulses that are much shorter than FEL coherence time. Thus, generation of attosecond pulses in XUV and soft X-ray regimes is enabled. 
Application of such a scheme to a user facility like FLASH would make it possible to create a unique source for attosecond science.

\section{Acknowledgment}

The author would like to thank Wim Leemans for his interest in this work, and Martin Dohlus for careful reading of the manuscript and useful suggestions.


\clearpage


\begin{thebibliography}{99}

\bibitem{attoscience}
P.B. Corkum and F. Krausz, Nat. Phys. 3(2007)381–7

\bibitem{attofel-oc}
E.L.~Saldin,
E.A.~Schneidmiller,
M.V.~Yurkov,
Optics Communications {\bf 212}(2002)377.

\bibitem{oc-2004-2}
E.L.~Saldin,~E. A.~Schneidmiller and M.~V.~Yurkov,
Optics Communications {\bf 237}(2004)153-164.

\bibitem{atto-b}
A.A.~Zholents and W.M.~Fawley,
Phys. Rev. Lett. {\bf 92}(2004)224801.

\bibitem{atto-e}
P.~Emma, Z.~Huang and M.~Borland,
Proc. of the FEL2004 Conference, p. 333, http://www.jacow.org

\bibitem{atto-f}
A.A.~Zholents and G.~Penn,
Phys. Rev. ST Accel. Beams {\bf 8}(2005)050704.

\bibitem{prstab-2006-2}
E.L.~Saldin, E.A.~Schneidmiller and M.V.~Yurkov,
Phys. Rev. ST Accel. Beams {\bf 9}(2006)050702.

\bibitem{ding}
Y.~Ding et al., Phys. Rev. ST Accel. Beams  {\bf 12}(2009)060703

\bibitem{huang-24as}
D.~Xiang, Z.~Huang, and G.~Stupakov,
Phys. Rev. ST Accel. Beams {\bf 12}(2009)060701

\bibitem{shuang}
S. Huang et al., Phys. Rev. Lett. {\bf 119}(2017)154801

\bibitem{duris}

J.~Duris et al., Nature Photonics {\bf 14}(2020)30

\bibitem{book}
E.L.~Saldin, E.A.~Schneidmiller and M.V.~Yurkov,
``The Physics of Free Electron Lasers'', Springer, Berlin, 1999

\bibitem{fawley}
W.M.~Fawley, Nucl. Instrum. and Methods {\bf A593}(2008)111

\bibitem{FL-CDR}
M. ~Beye, S.~Klumpp and W.~Wurth (Eds.), 
``FLASH2020+. Upgrade of FLASH. Conceptual Design Report'', DESY Hamburg, March 2020, 
DOI: 10.3204/PUBDB-2020-00465

\bibitem{rev-tap}
E.A.~Schneidmiller and M.V.~Yurkov,
Phys. Rev. ST-AB {\bf 16}(2013)110702

\bibitem{stupakov-taper}
Z.~Huang and G.~Stupakov,
Phys. Rev. ST-AB {\bf 8}(2005)040702

\bibitem{bon-rho}
R.~Bonifacio, C.~Pellegrini and L.M.~Narducci,
Opt. Commun. {\bf 50}(1984)373

\bibitem{bon-short}
R. Bonifacio, et al.
Phys. Rev. Lett. {\bf73}(1994)70

\bibitem{we-short}
E.L.~Saldin, E.A.~Schneidmiller and M.V.~Yurkov,
Nucl. Instrum. and Methods {\bf A507}(2003)101

\bibitem{bend-kulipanov}
G.N. Kulipanov, A.S. Sokolov and N.A. Vinokurov,
Nuclear Instruments and Methods in Physics Research {\bf A375}(1996)576 

\bibitem{macarthur}
J.~P.~MacArthur, 
A.~A.~Lutman, J.~Krzywinski, and Z.~Huang,
Physical Review X {\bf 8}(2018)041036

\bibitem{lutman-circ}
A.A.~Lutman et al., 
Nature Photonics {\bf 10}(2016)468

\bibitem{thompson}
N.R.~Thompson, 
Proc. of 39th Free Electron Laser Conf., 2019, Hamburg, Germany, p. 658; doi:10.18429/JACoW-FEL2019-THP033 

\bibitem{flash-nat-phot}
W.~Ackermann et al., Nature Photonics {\bf 1}(2007)336

\bibitem{njp}
K. Tiedtke et al., New Journal of Physics {\bf 11}(2009)023029

\bibitem{zemella}
J.~Zemella and M.~Vogt,
Proc. of 10th Int. Particle Accelerator Conf., 2019, Melbourne, Australia, p. 1744; doi:10.18429/JACoW-IPAC2019-TUPRB02

\bibitem{simplex}
T.~Tanaka,
J. Synchrotron Rad. {\bf 22}(2015)1319

\bibitem{duesterer}
I.J.B.~Macias et al.,
Optics Express {\bf 29}(2021)10491

\bibitem{sc-und}
G.~Geloni, E.L.~Saldin, E.A.~Schneidmiller and M.V.~Yurkov,
Nuclear Instruments and Methods in Physics Research {\bf A583}(2007)228 



\end{thebibliography}
\end{document}